\documentclass[aps,pra,showpacs,twocolumn,floatfix]{revtex4}
\usepackage{dcolumn} 
\usepackage{amsmath}
\usepackage{bm}
\begin{document}
\title{Off-diagonal hyperfine interaction between the $6p_{1/2}$ and $6p_{3/2}$
levels in $^{133}$Cs}
\author{W. R. Johnson}
\email{johnson@nd.edu} \homepage{www.nd.edu/~johnson}
\affiliation{Department of Physics, 225 Nieuwland Science Hall\\
University of Notre Dame, Notre Dame, IN  46556}
\author{H. C. Ho}
\email{hho1@nd.edu}
\affiliation{Department of Physics, 225 Nieuwland Science Hall\\
University of Notre Dame, Notre Dame, IN  46556}
\author{Carol E. Tanner}
\email{ctanner@nd.edu}
\affiliation{Department of Physics, 225 Nieuwland Science Hall\\
University of Notre Dame, Notre Dame, IN 46556}
\author{Andrei Derevianko}
\email{andrei@unr.edu} \homepage{http://www.physics.unr.edu/~tap}
\affiliation{Department of Physics, University of Nevada,
 Reno, NV 89557}

\begin{abstract}
The off-diagonal hyperfine interaction between the $6p_{1/2}$ and $6p_{3/2}$
states in $^{133}$Cs is evaluated in third-order MBPT giving
37.3 Hz and 48.3 Hz, respectively, for second-order energies of the $6p_{3/2}$ 
$F=3$ and $F=4$ levels.
This result is a factor of 10 smaller than one obtained from an uncorrelated
first-order Dirac-Hartree-Fock calculation and used in the analysis of a 
recent high-precision ($\leq 2$ kHz) measurement of the $6p_{3/2}$ hyperfine structure
[Gerginov {\it et al.} Phys.\ Rev.\ Lett.\ {\bf 91}, 72301 (2003)]. 
The factor of 10 difference has negligible effect on the conclusions of
the recent experiment but will become important for experiments carried out at 
a precision of better than 1 kHz.  
\end{abstract}
%
%

\date{\today}
%
%
\maketitle

\section{Introduction}

In the recent study of the hyperfine structure of the $6p_{3/2}$
state of $^{133}$Cs by \citet{GDT:03}, intervals between hyperfine 
levels were measured to an accuracy of $\leq 2$ kHz, which was sufficient 
to give, for the first time, a non-zero value for the $c$ hyperfine constant.
The value of the nuclear octupole moment
of $^{133}$Cs obtained from $c$ was $\Omega = 0.82(10)\, b\times \mu_N$, which is 
about a factor of 40 larger than 
nuclear shell-model prediction $\Omega_\text{n.s.m} = 0.022\, b\times \mu_N$, 
motivating a re-examination of corrections to the hyperfine constants.
One such correction is the second-order hyperfine interaction between the
$6p_{3/2}$ and $6p_{1/2}$ states. An estimate of this correction, based on an 
independent particle model of the cesium atom, was used in \cite{GDT:03}.
In the present work, we carry out a detailed third-order MBPT calculation
and obtain corrections to the $6p_{3/2}$ hyperfine levels
that are a factor of 10 smaller than the values used in \cite{GDT:03}. 
Revised values of the hyperfine constants $a$, $b$, and $c$,
obtained using the present results for the second-order hyperfine energies 
agree with those reported in \cite{GDT:03} to within the error estimates.      
However, for future experiments, especially experiments aimed at a precision of
better than 1 kHz, it will be important to use the correlated values of the
corrections presented here, rather than the larger values given in \cite{GDT:03}. 
   
\section{Perturbation Expansion}
We write the hyperfine interaction in the form
\[
H_{\rm hf} = \sum_{k\lambda}(-1)^\lambda T^{(k)}_{-\lambda}\, M^{(k)}_\lambda ,
\]
where $T^{(k)}_{-\lambda}$ is an irreducible tensor operator
acting in the electron sector and $M^{(k)}_\lambda $ is an
irreducible tensor operator acting in the nuclear sector. 
The first-order hyperfine correction to the energy of a state $|1\rangle$ is
\begin{multline}
\lefteqn{
W^{(1)}_F = \left< 1 |H_{\rm hf} | 1 \right> }\hspace{0em} \nonumber \\
 = \sum_k (-1)^{I+J+F}
\left\{
\begin{array}{ccc}
J   & I   & F \\
I   & J   & k
\end{array}
\right\}
\left< J \| T^{(k)} \| J \right>
\left< I \| M^{(k)} \| I \right> .
\end{multline}
The nuclear matrix elements are given in terms of
conventional nuclear moments through
\begin{eqnarray*}
 \left< II | M^{(1)}_0 | II \right> &=& \mu\  \\
 \left< II | M^{(2)}_0 | II \right> &=& \frac{1}{2} Q \\
 \left< II | M^{(3)}_0 | II \right> &=& -\Omega.
\end{eqnarray*}
Here, $\mu$ is the nuclear magnetic dipole moment, $Q$ is the nuclear electric
quadrupole moment,
and $\Omega$ is the nuclear magnetic octupole moment.
With these definitions, we introduce the conventional hyperfine constants 
$a$, $b$, and $c$ through the relations:
\begin{align}
a\ =\ & \frac{\mu}{IJ} \left< JJ | T^{(1)}_0 | JJ\right>\\
b\ =\ & 2 Q            \left< JJ | T^{(2)}_0 | JJ\right>\\
c\ =\ & -\Omega        \left< JJ | T^{(3)}_0 | JJ\right> ,
\end{align}
and write the first-order hyperfine energy as
\begin{multline}
W^{(1)}_F = \frac{1}{2}\, K\, a  
 + \frac{3K_+-4J_+I_+}{8I(2I-1)J(2J-1)}\, b\\
 + \frac{5K^2(K+4)-4K[3J_+I_+-J_+-I_+-3]-20J_+I_+}
     {I(2I-1)(2I-2)J(2J-1)(2J-2)}  \, c   , \label{w1}
\end{multline}
where 
$I_+ = I(I+1)$, 
$J_+ = J(J+1)$,
$F_+ = F(F+1)$,
$K = F_+-J_+-I_+$, and
$K_+ = K(K+1)$.

The second-order (in the hyperfine interaction) energy
of a state is given by
\[
W_F^{(2)} = \sum_{n\neq 1} \frac{\left< 1 |H_{\rm hf}| n\right>
\left< n |H_{\rm hf}| 1\right> }{E_1-E_n} .
\]
For the state $|1\rangle= |6p_{3/2}\rangle $ of Cs, the second-order hyperfine energy
is dominated by the single state $|n\rangle = |2\rangle = |6p_{1/2}\rangle$. 
Moreover, the largest
contribution from this state is the one associated with the magnetic dipole term
$k=k'=1$.  After angular reduction, the second-order energy is 
\begin{equation}
W^{(2)}_F = \left\{
\begin{array}{ccc}
J_2 & I   & F \\
I   & J_1  & 1
\end{array}
\right\} ^2
\frac{\left| \left< J_2 \| T^{(1)} \| J_1  \right>\right|^2
\left| \left< I \| M^{(1)} \| I \right>\right|^2}{E_1-E_2} . \label{eq2}
\end{equation}
Contributions to the second-order energy from the nuclear quadrupole
moment have been evaluated in lowest order and found to change the 
values obtained from Eq.~(\ref{eq2}) by less than 1\%. 
The fine-structure interval $E_1-E_2$ in the denominator
is determined as the difference between $f_{D2}$ the 
centroid of the $6s\ ^2 S _{1/2} \rightarrow 6p\ ^2 P _{3/2}$ transition
\cite{UDEM:00,GTD:04} and $f_{D1}$ the centroid of the 
$6s\ ^2 S _{1/2} \rightarrow 6p\ ^2 P_{1/2}$ transition \cite{UDEM:99},
both of which have been measured to high precision. One obtains
$E_1-E_2 = f_{D2}-f_{D1}= 1.6609669667(11)\times 10^7$ MHz. 
\section{Numerical Estimates}
Correlation corrections to hyperfine
matrix elements in alkali-metal atoms are large. Thus, for example, a lowest-order
Dirac-Hartree-Fock calculation of the hyperfine constant $a_{3/2}$ for the $6p_{3/2}$
state of Cs
(which is proportional to the diagonal matrix element 
$\left< {3/2} \| T^{(1)} \| {3/2}  \right>$)
leads to a result that is a factor of two smaller than the experimental
value. One expects (and indeed finds) corrections of a similar size 
for the off-diagonal
matrix element $\left< {1/2} \| T^{(1)} \| {3/2}  \right>$ 
appearing in the
numerator of the expression for the second-order hyperfine energy. If we
assume that the relative size of the correlation corrections to the two 
matrix elements mentioned above are the same, 
then we can determine the ratio  $\left< {1/2} \| T^{(1)} \| {3/2}\right>/
\left< {3/2} \| T^{(1)} \| {3/2}  \right>$
 by means of a lowest-order calculation
and, using that ratio together with the experimental value of $a_{{3/2}}$, 
obtain an accurate value for $\left< {1/2} \| T^{(1)} \| {3/2}  \right>$.
That was the strategy used to obtain the values 
$W_3^{(2)}=401$ Hz and $W_4^{(2)}=520$ Hz 
for the $6p_{3/2}$ state of $^{133}$Cs quoted in Ref.~\cite{GDT:03}. 
(The ratio of matrix elements was determined in the non-relativistic 
approximation and did not depend on details of the $6p$ wave function. 
The non-relativistic approximation is not a
serious problem, however, since the ratio obtained using relativistic Dirac-Hartree-Fock 
wave functions differs from the non-relativistic ratio by less than  5\%.)

\section{Correlation Corrections}
The estimates made in the previous section depend on the the assumption
that correlation corrections to reduced matrix elements of the hyperfine operator
$\left< j \| T^{(1)} \| j'  \right>$ are independent of the total
angular momentum $j$ of the $6p_j$ state. To test that assumption, we carried out
correlated third-order MBPT calculations of the three $6p_j$ matrix elements.

\begin{table}
\caption{MBPT contributions to the hyperfine constants $a_{1/2}$ and 
$a_{3/2}$ (MHz) of the
$6p_{1/2}$ and $6p_{3/2}$ states, respectively, and to the off-diagonal
matrix element $ g T^{(1)} = g_I\, \langle 1/2 \| T^{(1)} \| 3/2 \rangle$ (MHz)  are presented  .
The resulting third-order hyperfine constants are are compared with experiment. 
\label{tab1}}
\ruledtabular
\begin{tabular}{cddd}
\multicolumn{1}{c}{Term} &
\multicolumn{1}{c}{$a_{1/2}$} &
\multicolumn{1}{c}{$a_{3/2}$} &
\multicolumn{1}{c}{\small $g T^{(1)}$} \\
\hline
  1st        & 160.88&  23.92&   26.97\\
  2nd        &	40.66&	18.84&	-34.15\\[0.3pc]
  Bruck      &	84.40&	16.08&	 -1.12\\
  St Rad     &	 5.43&	-7.51&	 24.85\\
  Norm       &	-1.20&	-0.23&	  0.04\\[0.3pc]
  3rd        &	88.62&	 8.33&	 23.77\\
\cline{2-4}
  Total       & 290.17&	51.09&	 16.59\\
  Expt.       & 291.89& 50.29&  
\end{tabular}
\ruledtabular
\end{table}

In Table~\ref{tab1}, we give a detailed breakdown of contributions
to the third-order matrix elements. Formulas for the first-, second-, and
third-order matrix element are given in \cite{BGJS:87}. We use a
modified version of these formulas in which: (a) the sum of the
second-order
matrix element and the third-order contribution to the random-phase
approximation (RPA) is replaced by the exact solution to the RPA equations,
and (b) all one-electron matrix elements in third-order are replaced 
by their RPA counterparts. These modifications give dipole
transition matrix elements that are gauge invariant in second- and
third-order \cite{SJ:00}. The third-order hyperfine constants for
the $6p_{1/2}$ and $6p_{3/2}$ states evaluated in this way are within a few
percent of experiment. Since we use the same method to evaluate diagonal
and off-diagonal matrix elements, we expect the third order off-diagonal 
matrix element to be accurate to a few percent. 

Substituting the third-order off-diagonal matrix element given in 
Table~\ref{tab1} into Eq.~(\ref{eq2}), we find $W_3^{(2)} = 37.3$ Hz and 
$W_4^{(2)} = 48.3$ Hz for the $6p_{3/2}$ state. Combining the second-order
corrections with the
observed $6p_{3/2}$ hyperfine intervals (MHz) from \cite{GDT:03}
\begin{eqnarray*}
W_5-W_4 &=& 251.0916(20) \\ 
W_4-W_3 &=& 201.2871(11) \\ 
W_3-W_2 &=&  151.2247(16) ,
\end{eqnarray*}
we obtain the following values for the hyperfine constants (MHz)
\begin{eqnarray*} 
a&=& 50.28825(23)\\
b&=&-0.4940(17)\\
c&=&  0.00056(7) .
\end{eqnarray*}
These values agree within error limits with those found in \cite{GDT:03}.

\begin{acknowledgments}
The work of W.R.J. and H.C.H. was supported in part by National Science
Foundation (NSF) Grant No.\ PHY-01-39928.
The work of A.D. is supported in part by NSF 
Grant No.\ PHY-00-99419 and in part by a National Institute of Standards and Technology 
(NIST) precision measurement grant.  
The work of C.E.T. is supported in part by NSF
Grant No.\ PHY-99-87984 and in part by the U. S. Department of Energy (DOE) under Grant
No.\ DE-FG02-95ER14579.  
Any opinions, findings, and conclusions or recommendations 
expressed in this material are those of the authors and do not necessarily reflect 
the views of NSF, NIST, or DOE.
\end{acknowledgments} 


\end{document}